\documentclass[aps,prl,10pt,twocolumn,superscriptaddress,footinbib,flushbottom]{revtex4-1}%
\usepackage{amssymb,amsmath,amsfonts,bm,dcolumn}
\usepackage{graphicx,color}
\usepackage[squaren]{SIunits}
\usepackage[english]{babel}
\usepackage{tabularx}
\usepackage{tikz}

\renewcommand{\vec}[1]{\boldsymbol{\mathrm{#1}}}%
\newcommand{\uvec}[1]{\bm{\hat{\mathbf{#1}}}}
\newcommand{\abs}[1]{\lvert#1\rvert}%
\newcommand{\mean}[1]{\langle #1 \rangle}

\newcommand{\dif}{\mathrm{d}}
\newcommand{\del}[1]{\delta \hspace{-1pt} #1}
\newcommand{\Del}[1]{\Delta \hspace{-1pt} #1}
\newcommand{\integral}[4]{\int_{#1}^{#2} \! #3 \, \dif#4}%

\newcommand{\ii}{\mathrm{i}}
\renewcommand{\Re}{\operatorname{Re}}%

\newcommand{\DiracDelta}[1]{\operatorname{\delta}(#1)}

\newcommand{\ie}{i.e.}%

\usepackage{ulem}

\newcommand{\mode}{\tau}
\newcommand{\fct}[1]{#1}

\newmuskip\pFqmuskip
\newcommand*\pFq[6][8]{%
	\begingroup 
	\pFqmuskip=#1mu\relax
	\mathchardef\normalcomma=\mathcode`,
	\mathcode`\,=\string"8000
	\begingroup\lccode`\~=`\,
	\lowercase{\endgroup\let~}\pFqcomma
	{}_{#2}F_{#3}{\left[\genfrac..{0pt}{}{#4}{#5};#6\right]}%
	\endgroup
}
\newcommand{\pFqcomma}{{\normalcomma}\mskip\pFqmuskip}

\graphicspath{ {figure/} }

\begin{document}
	
	\title{Inertial self-propelled particles in anisotropic environments}
    
    \author{Alexander R. Sprenger}
	\affiliation{Institut f\"ur Theoretische Physik II: Weiche Materie, Heinrich-Heine-Universit\"at D\"usseldorf, D-40225 D\"usseldorf, Germany}
	
	\author{Christian Scholz}
	\affiliation{Institut f\"ur Theoretische Physik II: Weiche Materie, Heinrich-Heine-Universit\"at D\"usseldorf, D-40225 D\"usseldorf, Germany}
	
	\author{Anton Ldov}
	\affiliation{Institut f\"ur Theoretische Physik II: Weiche Materie, Heinrich-Heine-Universit\"at D\"usseldorf, D-40225 D\"usseldorf, Germany}
	
	\author{Raphael Wittkowski}
	\affiliation{Institut f\"ur Theoretische Physik, Center for Soft Nanoscience, Westf\"alische Wilhelms-Universit\"at M\"unster, D-48149 M\"unster, Germany}
	
	\author{Hartmut L\"owen}
	\affiliation{Institut f\"ur Theoretische Physik II: Weiche Materie, Heinrich-Heine-Universit\"at D\"usseldorf, D-40225 D\"usseldorf, Germany} 

	\date{\today}
	
	\begin{abstract}
        \section{Abstract}
		Self-propelled particles in anisotropic environments can exhibit a motility that depends on their orientation. 
		This dependence is relevant for a plethora of living organisms but difficult to study in controlled environments. 
		Here, we present a macroscopic system of self-propelled vibrated granular particles on a striated substrate that displays orientation-dependent motility. 
		An extension of the active Brownian motion model involving orientation-dependent motility and inertial effects reproduces and explains our experimental observations. 
		The model can be applied to general $n$-fold symmetric anisotropy and can be helpful for predictive optimization of the dynamics of active matter in complex environments.
	\end{abstract}
	
	\pacs{???}
	
	\maketitle

    \section{Introduction}
	
	The survival of organisms in complex environments essentially depends on their fitness and strategy to react and adapt to external conditions. In particular, a realistic  environment is never isotropic but typically anisotropic, i.e., its traversability depends on the direction of motion \cite{Aranson2018}. 
	Anisotropy can be caused on various scales by many different means: by an external force arising from gravity \cite{EnculescuS2011,HaederH2017}, viscosity \cite{LiebchenMtHL2018}, light \cite{LozanotHLB2016}, and chemical gradients \cite{HongBKSV2007}, electromagnetic fields \cite{SprengerFRAIWL2020}, through steric confinement by channels, veins, and anisotropic porous media \cite{WysockiEG2015,ParisiHZ2018}, or by motion in a liquid-crystalline \cite{MushenheimTTWA2014,ZhouSLA2014,GuillamatIJS2016,TonerLW2016,PengTGWL2016} or crystalline \cite{VolpeBVKB2011,BrownVDVLP2016,vanderMeerFD2016} medium. 
	Anisotropic environments can have a pronounced impact on the motion of self-propelled particles. 
	These ``active'' particles convert energy from their environment into directed motion and comprise both living organisms and artificial inanimate objects, like activated colloids \cite{RomanczukBELSG2012,ElgetiWG2015,Menzel2015}, granules \cite{JaegerNB1996,AltshulerPGZM2013,DauchotD2019,ScholzLPEL2021,TorresMenendezABP2022}, and robots \cite{LeymanOWV2018,ZionBD2021,FalkAJM2021}. 
	Standard models of self-propelled particles \cite{BechingerDLLRVV2016} assume that the propulsion force is isotropic in the sense that it always points into the direction of the particle orientation with a constant self-propulsion speed even in an inhomogeneous environment \cite{GoshLMN2015,MagieraB2015,Grossmann2015,GeiselerHMMS2016,GeiselerHM2017a,GeiselerHM2017b,SharmaB2017}. 
	In anisotropic environments, a dependence of the self-propulsion speed of the particle on its orientation is frequently observed, \ie, some biological organisms react to their environment in a sense that the propulsion force depends on their orientation relative to the environment. For instance, microorganisms can  move faster towards light sources \cite{BennettG2015} or in the direction of food sources \cite{Cates2012}. 
	Additionally, flying animals such as bees and birds control their flying speed by relative changes of their environment, which in turn leads to anisotropic flying velocities within structured environments \cite{BairdKWWD2011,ScholtyssekDKB2014,BhagavatulaCIS2011}.
	Similarly, anisotropic movement is also observed for smaller insects like ants in guiding structures \cite{BolekW2015,FeinermanPGFG2018}. 
	Those macroscopic self-propelled particles in low-friction environments (e.g., such as flying insects) where the effect of anisotropy is most prominent, are also governed by inertial effects \cite{DevereuxTTT2021}. This poses a challenging problem because inertia introduces correlations that can persist for longer times \cite{GhoshLMM2015,WalshWSOBM2017,ScholzJLL2018,Loewen2020,CapriniM2020,GutierrezMartinezS2020,HerreraS2021,SprengerJIL2021,NguyenWL2022,SprengerCLW2023}.
 
	In this communication, we present an experimental realization of a self-propelled granular particle on an anisotropically structured substrate, which exhibits orientation-dependent motility. 
	We observe pronounced anisotropy in the motion of the particle, which is well explained analytically by an extension of the active Brownian motion model with inertia and orientation-dependent motility. The orientation-dependence can be written in terms of a Fourier series which allows a general solution for anisotropic motility that can be applied to our experiments. 
    Our findings establish a class of active matter models useful for anisotropic environments and shed light on the potential self-propulsion strategies of organisms in such anisotropies.
	The analytical results of our model can be particularly useful for predictive optimization of control parameters of artificial active agents, such as robots \cite{LeymanOWV2018,ZionBD2021,FalkAJM2021}, to better explore anisotropic environments \cite{VolpeV2017}.

\section{Results}
	
\subsection{Experimental observation of anisotropic self-propulsion}

	\begin{figure*}
		\includegraphics[width=16cm]{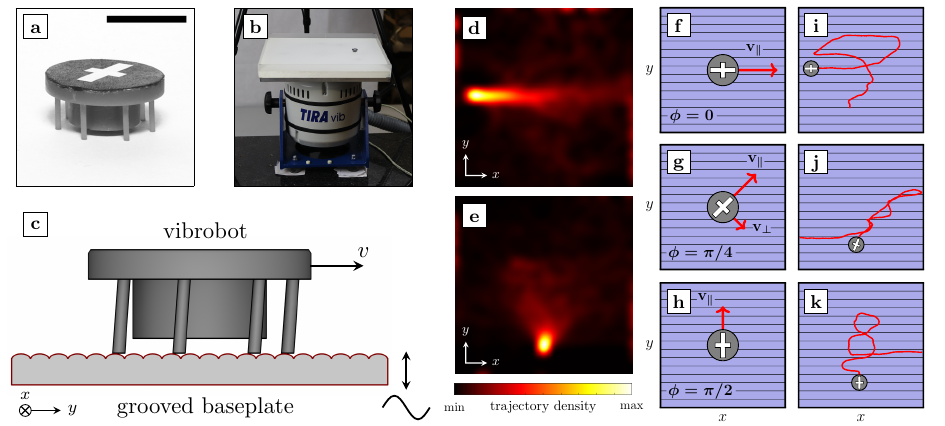}
		\caption{Description of experimental system and observations. 
		\textbf{a} Vibrationally driven self-propelled particle (vibrobot) manufactured by 3D printing. The white cross indicates the particle orientation. Scale bar represents 1\,cm. \textbf{b} Experimental setup: Rectangular acrylic baseplate attached to an electromagnetic shaker. The size (width~$\times$~length) of the top-mounted plate equals 30\,cm~$\times$~30\,cm. \textbf{c} Cross-section of the anisotropic substrate (lenticular foil) with particle to scale.
		\textbf{d}, \textbf{e} Trajectory density for vibrobots starting parallel (\textbf{d}) and perpendicular (\textbf{e}) to grooves with an excitation amplitude $A$ = 1.28\,g. \textbf{f}-\textbf{h} Sketch of the two velocity contributions. The particle moves with increased velocity $\text{v}_\parallel$ when aligned along the grooves (\textbf{f}). When orientated diagonally, the particle moves with average velocity $\text{v}_\parallel$  along its orientation while simultaneously experiencing active propulsion $\text{v}_\perp$ perpendicular to it (\textbf{g}). The particle moves with decreased velocity $\text{v}_\parallel$  when perpendicularly aligned to the grooves (\textbf{h}). \textbf{i}-\textbf{k} Three representative trajectories with an excitation amplitude $A$ = 1.60\,g. The persistence length is noticeably shorter for perpendicularly aligned particles than for parallel aligned particles. Length ratios and velocity contributions are not to scale.
		}
		\label{fig:experiment}
	\end{figure*}
 
Macroscopic active matter with orientation-dependent motility can be realized from self-propelled 3D-printed agents called vibrobots (see Fig.~\ref{fig:experiment}a) on structured substrates. These particles are excited by vertical vibrations generated by a rectangular acrylic baseplate attached to an electromagnetic shaker. 
The particles stand on slightly tilted legs, which causes the particles to hop forward.
These legs are all tilted equally along the orientation (or symmetry axis) of the particle.
The baseplate is covered with a lenticular plastic sheet on top, which is the source for the anisotropic motility.
The experimental setup is depicted in Fig.~\ref{fig:experiment}b.
An illustration with a side-view of the particle resting on such a grooved surface are shown in Fig.~\ref{fig:experiment}c.
The vibration frequency is set to $f=80\,\mathrm{Hz}$. In this frequency range, the plate vibration is sufficiently uniform \cite{ScholzJLL2018}.
Three different peak acceleration amplitudes $A = 1.28\,g$, $1.44\,g$, and $1.60\,g$ are investigated, which varies the motility and motion properties of the vibrobot. For this choice of $f$ and $A$, the vibration is strong enough to ensure stable vibrobot motion, but not too strong to prevent particles from falling over.

We find pronounced anisotropy in the motion of the particle and observe a modulation of the velocity parallel but also perpendicular to the orientation of the grooves, as well as an increased activity with increasing excitation amplitude. The motion of the particles is illustrated in Supplementary videos 1 - 6, where we show a montage of all measured trajectories for each excitation amplitude as well as for parallel and perpendicular initial orientation, respectively. From the trajectories, the anisotropy is already visible by the naked eye, in particular when comparing parallel and perpendicular starting orientations.

This anisotropy is best illustrated when displaying all recorded trajectories (integrated and smoothed) and distinguishing parallel and perpendicular initial orientations, as shown in Fig.~\ref{fig:experiment}d, e.
For particles starting parallel to the grooves, we observe that the peak of the density (which is linked to the starting position of the particles) is broad along and narrow perpendicular to the starting orientation since the particles tend to move faster parallel to the grooves and therefore propagate further before they reorient. 
In the case of perpendicular starting orientation, the density spreads more around the peak, since particles reorient near to the starting position. 
Hence the persistence length depends on the orientation of the particle. Surprisingly, from individual particle trajectories, we also identify a driving-force component perpendicular to the orientation, whenever a particle is not moving exactly parallel or perpendicular to the grooves. 
	
The anisotropic self-propulsion is caused by the grooved surface of the vibrating plate.
Our conjecture is that this is due to the strong dependence of the particle speed on the relative inclination angle between legs and surface \cite{KoumakisGMPL2016}. 
When resting on the vibrating plate, the legs are bent along the orientation of the particle. This deformation stores elastic energy. Then, after detaching from the base, the energy is released and the vibrobot jumps forward. 
When the particle is oriented perpendicular to the grooves, the legs face an elliptical half-cylinder and the relative inclination angle of the legs is decreased (see Fig.~\ref{fig:experiment}c). As a result, the legs will bend less compared to the case where the particle is oriented along the grooves. If the particle is diagonally aligned with the grooves, the legs will not bend along the orientation and the particle experiences a force perpendicular to its orientation. This in fact results in propulsion perpendicular to the orientation of the particle.
In Fig.~\ref{fig:experiment}f-h, we illustrate the two velocity contributions for three different orientations of the particle. 
	
As described in the literature, we also observe orientational fluctuations, caused by an instability of the driving mechanism to the microscopic surface roughness, and inertial delay effects due to the mass of the particles \cite{ScholzJLL2018, Leoni2020surfing}. When vibrobots are excited above a certain amplitude threshold, they begin to tumble \cite{ScholzP2016}. As a result, they randomly reorient while moving and eventually change the direction of their path. Figure~\ref{fig:experiment}i-k shows three representative trajectories with different initial orientations.
Clearly, the particle does not show a deterministic motion, apart from short-time correlations due to initial orientation and inertia. The particle rather undergoes an anisotropic two dimensional random walk with a certain persistence length. 

Due to the simplicity of our particles, compared to living active matter, our experiment allows us to investigate kinetic properties of particles with orientation-dependent motility, which can be useful for optimization of motion and search strategies of active matter in general. This requires an analytical description of the motion that captures the essential properties of the particle and must be applicable to general cases of anisotropic motility.

\subsection{Langevin dynamics model}

Finding an analytical description for macroscopic self-propelled systems can be challenging due to the complex interaction of particles and environment. 
Here, we model those interactions with an effective driving force and thereby introduce a minimal model, where the interplay of orientation-dependent motility, inertia, and fluctuations, is treated in terms of a generalized active Langevin dynamics model.
Our model reproduces the experimental observations quantitatively despite its complex anisotropic nature.  

We assume that the particle has non-negligible mass $M$ and moment of inertia $J$. The motion of such an underdamped particle is in general characterized by the translational center-of-mass velocity $\dot{\vec{r}}(t)$ with the center-of-mass position $\vec{r}(t)$ and the time variable $t$ as well as by the angular velocity $\dot{\phi}(t)$ and the angle of orientation $\phi(t)$, which denotes the angle between the orientation vector $\uvec{n} = (\cos\phi,\sin\phi)$ and the positive $x$-axis.
By taking the above considerations into account, the translational and rotational motion of the particle is governed by the force balance between inertial, frictional, self-propulsive driving, and random forces and torques
{\allowdisplaybreaks
\begin{gather}%
	M \, \ddot{\vec{r}}(t) + \gamma_\text{t} \, \dot{\vec{r}}(t) =  \gamma_\text{t} \, \vec{v}\big(\phi(t)\big) +  \sqrt{2 D_\text{t}} \, \gamma_\text{t} \, \vec{\xi}(t), 
	\label{eq:langevin_r}\\%
	J \, \ddot{\phi}(t) + \gamma_\text{r} \, \dot{\phi}(t)  = \gamma_\text{r} \, \omega + \sqrt{2 D_\text{r} } \, \gamma_\text{r} \, \eta(t).
	\label{eq:langevin_phi}%
\end{gather}%
}%
Here, $\gamma_\text{t}$ and $\gamma_\text{r}$ denote the translational and rotational friction coefficients, respectively. To take translational and rotational diffusion into account, the Langevin equations contain independent Gaussian white noise terms $\vec{\xi}(t)$ and $\eta(t)$, with zero means $ \mean{\vec{\xi}(t)} = \vec{0}$ and $ \mean{\eta(t)} = 0$ and delta-correlated variances $\mean{ \xi_{i} (t_{1}) \xi_{j}(t_{2})} = \delta_{ij} \DiracDelta{t_{1} - t_{2} }$ and $\mean{ \eta(t_{1}) \eta(t_{2})} = \DiracDelta{t_{1} - t_{2} } $, where $i, j \in \{x, y\}$. Therein, $D_\text{t}$ and $D_\text{r}$ are the translational and rotational short-time diffusion coefficients of the particle, respectively. The brackets $\mean{\dots}$ denote the noise average in the stationary state (meaning after losing correlation with initial conditions \cite{SprengerJIL2021}) and $\delta_{ij}$ is the Kronecker delta.

Most importantly, $\vec{v}(\phi)$ denotes an arbitrary orientation-dependent motility which accounts for the interaction between the particle and environment. 
For mathematical convenience, we represent $\vec{v}(\phi)$ as a Fourier series 
\begin{equation}%
\vec{v}(\phi) = \sum_{\substack{k=-\infty \\ k\neq 0}}^{\infty} \vec{c}_{k} \exp ( \ii k \phi ),
\label{eq:velocity}%
\end{equation}%
where $\vec{c}_{k}$ is the Fourier-coefficient vector of the mode $k$, and $\ii$ denotes the imaginary unit. 
This representation lets us solve the model for any type of orientation-dependence and then apply the results to our experimental system. 
In particular, this description can be used for different experimental realizations ranging from anisotropic illuminated Janus particles, triangular microparticles in traveling ultrasound waves, and the motion of living insects in guiding structures to the specific setup studied in this communication \cite{Uspal2019,VossW2022}. 
In general, for a given propulsion velocity $\vec{v}(\phi)$, these Fourier coefficients can be calculated as $\vec{c}_{k} =  \integral{-\pi}{\pi}{(\vec{v}(\phi)/(2 \pi)) \exp(- \ii k \phi)}{\phi}$ (thus we have after complex conjugation $\vec{c}_{k}^{*} = \vec{c}_{-k}$). 
The seminal case of isotropic propulsion is recovered for the two non-zero coefficients $\vec{c}_{\pm1} = v (1, \mp \ii)/2$.
Note that we exclude the mode $k=0$ in Eq.~\eqref{eq:velocity}, which would correspond to a drift velocity induced by a constant external force (e.g., gravity) not measured in the experiment. 

Moreover, as typical 3D-printed particles are not perfectly symmetrical, they tend to perform circular motions on long time scales. To capture this behaviour, we assume a systematic torque which acts on the particle and leads to an angular speed $\omega$. In contrast to $\vec{v}(\phi)$, we measured no orientational dependency in the angular speed which could in principle be caused by the anisotropic substrate. 

Concluding, our theoretical model depends on a number of parameters: the angular velocity $\omega$, the rotational diffusion coefficient $D_\text{r}$, the rotational friction time $\tau_\text{r} = J / \gamma_\text{r}$, the set of Fourier coefficients $\{ \vec{c}_k \}$ describing the anisotropic motility, the translational diffusion coefficient $D_\text{t}$ and the translational friction time $\tau_\text{t} = M / \gamma_\text{t}$. 
In the context of the experimental observations, we assume that the vibrobot is moving with an orientation-dependent velocity
\begin{equation}
	\vec{v}(\phi) = \left( \text{v}_{\parallel} + \del{\text{v}}_{\parallel} \cos(2 \phi) \right) \uvec{n}(\phi) - \del{\text{v}}_{\perp} \sin(2 \phi) \uvec{n}_{\perp}(\phi), 
	\label{eq:velocity_experimental}%
\end{equation}
where $\uvec{n}(\phi)=(\cos\phi,\sin\phi)$ is pointing parallel and $\uvec{n}_\perp(\phi)=(-\sin\phi,\cos\phi)$ is pointing perpendicular to the particle's orientation. 
The sine and cosine terms in Eq.~\eqref{eq:velocity_experimental} reflect the orientation dependence of the particle velocity and the symmetry of the system. 
This adds the parallel speed $\text{v}_{\parallel}$, the parallel speed anisotropy $\del{\text{v}}_{\parallel}$, and the perpendicular speed anisotropy $\del{\text{v}}_{\perp}$, leading to a total of 8 independent parameters. 
The four non-zero Fourier coefficients of Eq.~\eqref{eq:velocity_experimental} read $\vec{c}_{\pm1} = \text{v}_{\parallel} (1, \mp \ii)/2 + (\del{\text{v}}_{\parallel}+\del{\text{v}}_{\perp}) (1, \pm \ii)/4$ and $\vec{c}_{\pm3} = (\del{\text{v}}_{\parallel}-\del{\text{v}}_{\perp}) (1, \mp \ii)/4$.

These parameters are determined from analytic fits to the experimental results. 
We use temporal correlation functions, like the orientational correlation function $C(t) = \mean{\uvec{n}(t)\cdot \uvec{n}(0)}$ and the velocity correlation function $Z(t) = \mean{\dot{\vec{r}}(t) \cdot \dot{\vec{r}}(0)}$, to determine the relevant timescales and diffusion coefficients. Further stationary observables, like the mean translational velocity $\vec{v}_{0} = \mean{\dot{\vec{r}}(0)}$ and the mean angular velocity $\mean{\dot{\phi}(0)}$, are used to estimate all motility parameters. 
More information on the parameter estimation can be found in the Methods section and the parameter values are listed in Tab.~\ref{table:parameter}.
In the following, we compare the experimental data with analytic predictions derived from the theoretical model and discuss the anisotropy found in several observables.

\subsection{Comparison between analytical results and experiment}

\begin{figure}
	\includegraphics[width=\columnwidth]{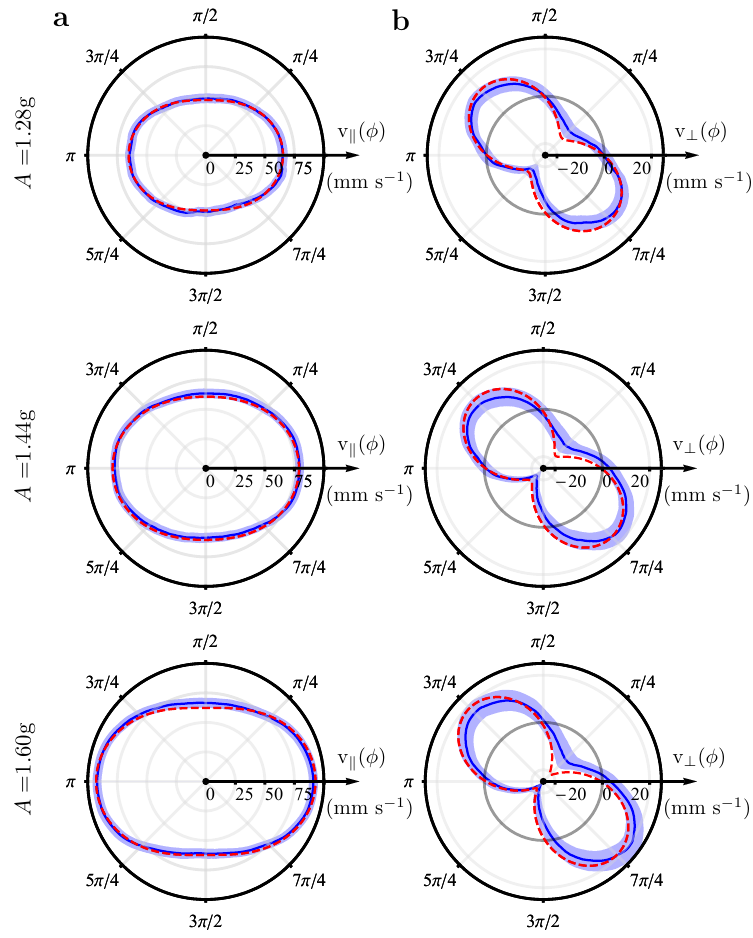}
	\caption{Orientation dependence of stationary velocity. \textbf{a} Stationary parallel velocity $\text{v}_\parallel$ and \textbf{b} stationary perpendicular velocity $\text{v}_\perp$ plotted as a function of the orientation angle $\phi$ for three different excitation amplitudes A = 1.28 g (upper row), A = 1.44 g (middle row), and 1.60 g (lower row). Solid dark blue and dashed red curves show the experimental data and analytical results, respectively. Blue experimental error intervals represent the standard error of the mean.}
	\label{fig:stationary_velocity}
\end{figure}

As described above, the mean self-propulsion strongly depends on the relative orientation of the particle with respect to the groove direction. The model describes this via two orthogonal velocity components.
In Fig.~\ref{fig:stationary_velocity}, we separately show the mean velocity along the body-axis $\text{v}_\parallel = \vec{v}_{0} \cdot \uvec{n}$ and perpendicular to it $\text{v}_\perp = \vec{v}_{0} \cdot \uvec{n}_\perp$ as  functions of the orientation $\phi$. 
The parallel contribution $\fct{v}_\parallel$ in Fig.~\ref{fig:stationary_velocity}a shows considerably greater propulsion along the grooves than perpendicular to them. 
For the perpendicular contribution (see Fig.~\ref{fig:stationary_velocity}b) we find the assumed $\sin(2\phi)$-modulation (see Eq.~\eqref{eq:velocity_experimental}), which has an alignment effect on the overall velocity direction in favor of the groove direction.
Overall, we measure increased activity for larger excitation amplitudes while the degree of anisotropy remains almost the same for all three measurements.  
From the theoretical side, the mean instantaneous velocity $\vec{v}_{0} = \mean{\dot{\vec{r}}(0)}$ at a specific orientation $\phi_{0}$ can be computed in general as instatanteous 
\begin{equation}
	\vec{v}_{0} = \frac{\tau_\text{r}}{\tau_\text{t}} \sum_{\substack{k=-\infty \\ k\neq 0}}^{\infty}  \vec{c}_{k}  e^{ \text{S}_{k} }  \text{S}_{k}^{-\Omega_{k}^{+} } \Gamma (\Omega_{k}^{+},0, \text{S}_{k})  e^{\ii k \phi_{0} },
	\label{eq:stationary_velocity}%
\end{equation}
with the dimensionless coefficients $\text{S}_k =  D_\text{r} \tau_\text{r} k^2$, $\Omega_{k}^{+} = D_\text{r} \tau_\text{r} k^2 +\ii \omega \tau_\text{r} k + \tau_\text{r} / \tau_\text{t}$,   and the generalized incomplete gamma function  $\Gamma(s,x_{1},x_{2}) = \integral{x_{1}}{x_{2}}{ t^{s-1} e^{-t}}{t}$.
The analytic result is plotted in Fig.~\ref{fig:stationary_velocity} and yields good agreement with the experimental data.
In contrast to overdamped motion, where the particle's mean velocity is simply equal to the internal self-propulsion velocity, here the particle moves on average with a smaller velocity due to inertial delay effects, \ie, $\abs{\vec{v}_0(\phi)} \le \abs{\vec{v}(\phi)}$.
Further, the faster varying contributions (i.e., the higher Fourier modes) of the propulsion are more affected by these inertial delay effects, resulting in a more isotropic mean velocity for increasing mass $M$. 
Conversely, the anisotropy is restored for increasing moment of inertia: $\lim_{J \to \infty}\vec{v}_0(\phi) = \vec{v}(\phi)$.

\begin{figure}
	\includegraphics[width=\columnwidth]{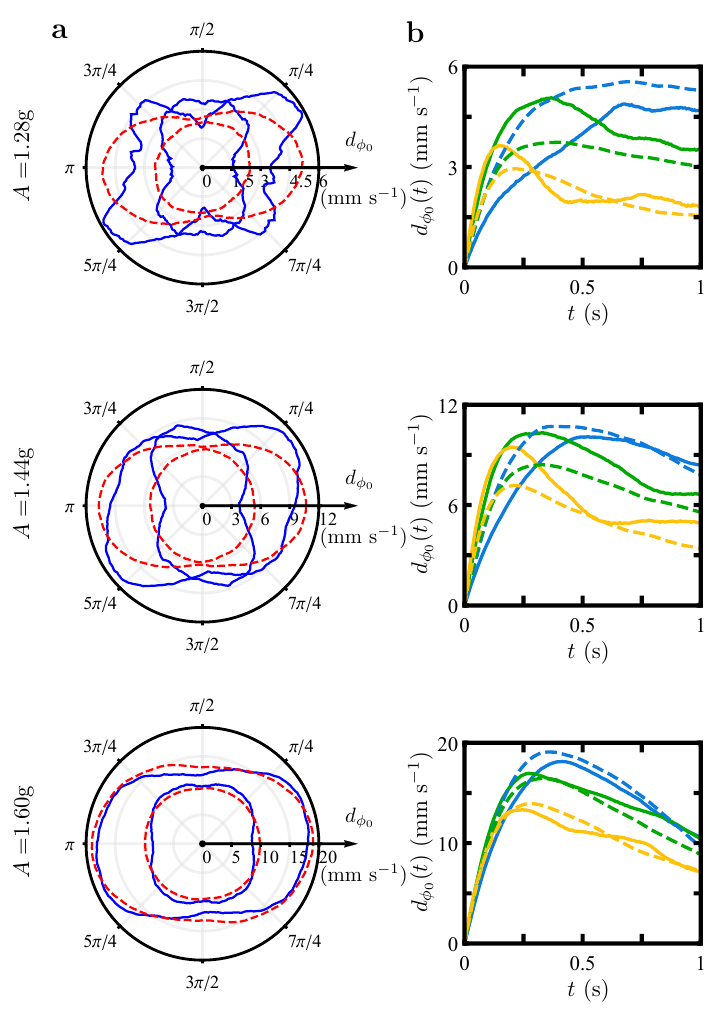}
	\caption{
	Anisotropic delay function. \textbf{a} The anisotropic delay function $d_{\phi_0}(t)$ plotted as function of the initial orientation $\phi_0$ after fixed times  $t = 0.1$\,s, $t = 0.4$\,s. 
	Solid blue and dashed red curves show the experimental and simulated data, respectively.
	\textbf{b} The anisotropic delay function $d_{\phi_0}(t)$ plotted as a function time $t$ for parallel  $\phi_0 = 0$  (cyan), diagonal $\phi_0 = \pi/4$ (green), and  perpendicular $\phi_0 = \pi/2$ (yellow) orientations, each. 
	Both for excitation amplitude $A$ = 1.28\,g (upper row), $A$ = 1.44\,g (middle row) and $A$ = 1.60\,g (lower row).
	Solid and dashed curves correspond to the experimental and simulated data (using the parameter values given in Tab. \ref{table:parameter}), respectively.}
	\label{fig:anisotropic_delay}
\end{figure}

\begin{figure}
	\includegraphics[width=\columnwidth]{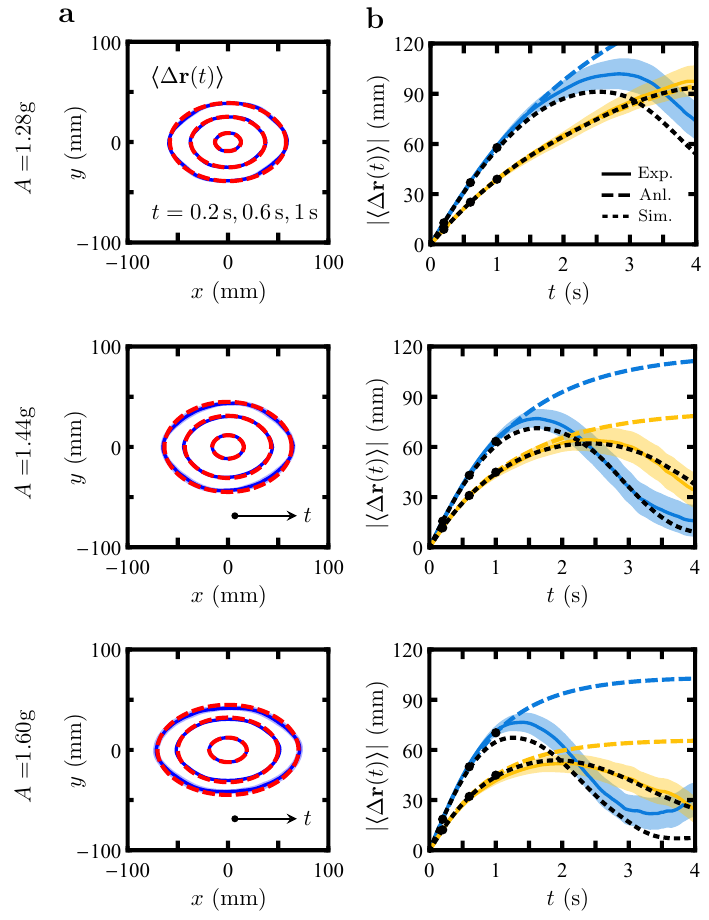}
	\caption{Mean displacement. Comparison between model and measurement with excitation amplitude $A$ = 1.28\,g (upper row), $A$ = 1.44\,g (middle row), and $A$ = 1.60\,g (lower row). \textbf{a} The anisotropic motion of the particle is visualized by plotting the mean displacement $\left\langle\Delta \vec{r}(\phi_0)\right\rangle$ for $\phi_0 \in [0,2\pi)$ and fixed times $t = 0.2$\,s, $t = 0.6$\,s and $t = 1.0$\,s. Solid blue and dashed red curves show the experimental data and analytical results, respectively. Light blue area expresses the standard error of the mean. \textbf{b} The absolute mean displacement $|\langle\Delta \vec{r}(t)\rangle|$ is plotted as a function of time $t$ for initial orientations $\phi_0 = 0$ (cyan) and $\phi_0 = \pi/2$ (yellow). Solid colored curves represent the experimental data and dashed colored curves the analytic results. In addition, dashed black curves depict simulation data for a particle in confinement.
	Black dots correspond to the experimental values for the fixed times of Fig.~\ref{fig:mean_displacement}a.
	Theoretical predictions and simulations use the parameters given in Tab.~\ref{table:parameter}.}
	\label{fig:mean_displacement}
\end{figure}

A suitable quantifier for the presence of inertial effects is the delay function
$\fct{d}(t) = \mean{\dot{\vec{r}}(t)\cdot\uvec{n}(0)} - \mean{\dot{\vec{r}}(0)\cdot\uvec{n}(t)}$ \cite{ScholzJLL2018,Loewen2020,SprengerCLW2023}. 
This function quantifies the average difference between the projection of the initial velocity on the orientation and the projection of the initial orientation on the velocity. In overdamped systems, this function is zero at all times. Here, we find that this function is significantly different from zero in particular for large excitation amplitudes $A$ (see the Methods section).
The standard delay function can be generalized to resolve anisotropy in the system by conditioning the average with a specific initial orientation $\phi_0$ at time $t=0$. 
In Fig.~\ref{fig:anisotropic_delay} we plot the anisotropic delay function $\fct{d}_{\phi_0}(t)$ both as a function of $\phi_0$ for given $t$ and as a function of $t$ for given $\phi_0$.
We compare the experimental data with simulations which follow Eqs.~\eqref{eq:langevin_r} and \eqref{eq:langevin_phi} and are initialized similar to the experiments.
The delay function is a highly fluctuating quantity making the experimental data difficult to interpret.  
The simulated data suggests an isotropic delay for short times and a larger delay along the grooves as time proceeds mimicking the modulation of the self-propulsion velocity. 
The simulated data always fits within the standard error of the experimental data.

For stochastic processes, it is common to analyze the first and seconds moments of the motion, i.e., the mean and mean square displacement.
In anisotropic systems, these quantities will strongly depend on the initial orientation of a particle. 
In Fig.~\ref{fig:mean_displacement}, we compare the experimental mean displacement $\mean{\Delta \vec{r}(t)}$ conditioned at different initial orientations $\phi_{0}$ with that resulting from our theoretical model. 
To demonstrate the effect of the orientation-dependent motility, we show the mean displacement as a function of the initial orientation $\phi_{0}$ after fixed times $t$ forming elliptic-like shapes in the $xy-$plane (see Fig.~\ref{fig:mean_displacement}a).
In Fig.~\ref{fig:mean_displacement}b, we plot the absolute mean displacement $|\langle\Delta \vec{r}(t)\rangle|$ as a function of time $t$ for particles which are initially orientated along the grooves (blue) and for those starting perpendicular to the grooves (red). 
The experimental data fit within theoretical results for short time, where the particle moves linearly in time with $\mean{ \Del{\vec{r}}(t)  } = \vec{v}_{0} t + \mathcal{O}(t^2)$. 
For longer time, confinement effects play an increasing role. 
Since recordings are stopped once a particle hits the boundary, events where the particle reorients beforehand dominate the statistic.
As a consequence, the measured mean displacement decreases for times larger than the mean first-passage time of hitting the boundary.
We perform simulations with absorbing boundaries and find an excellent agreement for all experimental accessible time scales (indicated by the black dashed curves in Fig.~\ref{fig:mean_displacement}b). 
Without confinement, the theoretical mean displacement saturates to an anisotropic persistence length $\vec{L}_\text{p} = \lim_{t \to \infty} \mean{\Delta \vec{r}(t)}$ for long times
\begin{equation}
	\vec{L}_\text{p} =  \vec{v}_{0} \tau_\text{t} +  \sum_{\substack{k=-\infty \\ k\neq 0}}^{\infty} \! \vec{c}_{k} \, \mode_{k} \, e^{\ii k \phi_{0}}, 
	\label{eq:persistence_length}%
\end{equation}
with the persistence time of mode $k$
\begin{equation}
	\mode_{k} = \tau_\text{r} e^{ \text{S}_{k} }  \text{S}_{k}^{-\Omega_{k}} \Gamma(\Omega_{k}, 0,\text{S}_{k})
	\label{eq:persistence times}%
\end{equation}
and $\Omega_{k} = D_\text{r} \tau_\text{r} k^2 + \ii \omega \tau_\text{r} k$.
The persistence length $\vec{L}_\text{p} $ consists of two contributions: the first term is given by the mean stationary velocity $\vec{v}_{0}$ which is damped over the translational friction time $\tau_\text{t}$.
The second term  in Eq.~\eqref{eq:persistence_length} describes the active propulsion getting decorrelated due to the rotational noise $D_\text{r}$. 
Again, the degree of anisotropy increases as a function of the moment of inertia $J$.
For vanishing angular speed $\omega=0$, we find the following asymptotic behavior for small and large $J$, respectively:
\begin{equation} \label{eq:persistence times_asymptotic}
	\mode_{k} \sim
	\begin{cases}
		\frac{1}{D_\text{r} k^2} \left( 1 + \frac{D_\text{r} k^2}{\gamma_\text{r}} J \right), & \text{for}~J \to 0, \\
		\frac{1}{k} \sqrt{\frac{\pi}{2 D_\text{r} \gamma_\text{r}}}\sqrt{J},  & \text{for}~J \to \infty.
	\end{cases}
\end{equation}

\begin{figure}
	\includegraphics[width=\columnwidth]{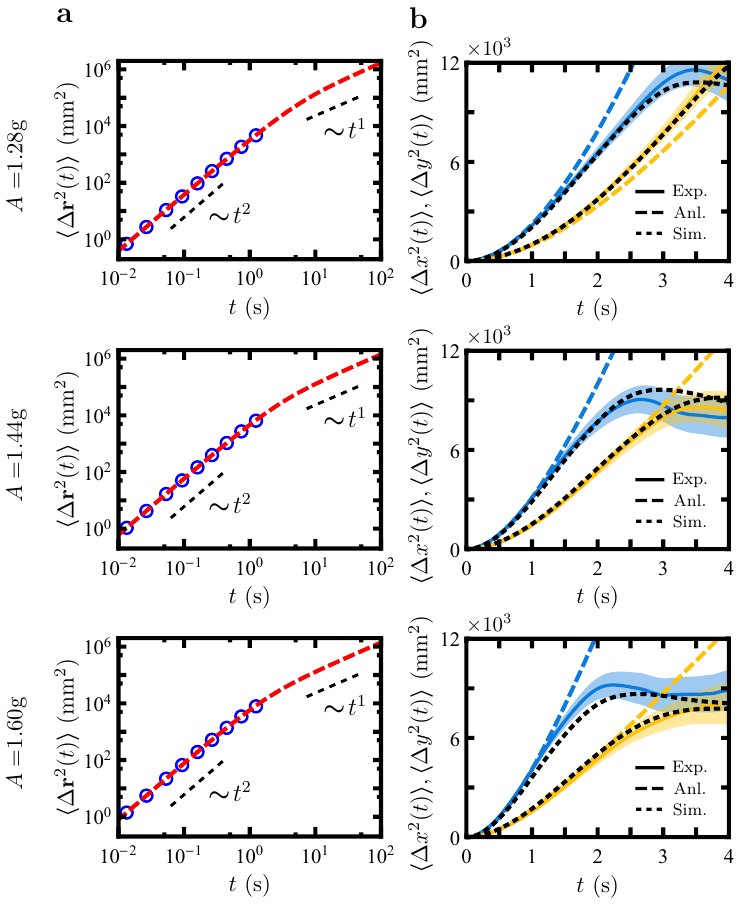}
	\caption{Mean square displacement. Comparison between model and measurement with excitation amplitude $A$ = 1.28\,g (upper row), $A$ = 1.44\,g (middle row), and $A$ = 1.60\,g (lower row).
	\textbf{a} The total mean-square displacement as a function of time $t$ (double logarithmic scaling). Open blue circles and dashed red curves show the experimental data and analytical results, respectively.
	\textbf{b} The mean-square displacement along the $x$-axis (cyan) and $y$-axis (yellow) as functions of time $t$. 
	Solid colored curves and dashed colored curves show the experimental data and analytical results, respectively. 
	Light colored areas represent the standard error of the mean.
	Dashed black curves show simulation data for a particle in confinement.
	Theoretical predictions correspond to the parameters given in Tab.~\ref{table:parameter}.}
	\label{fig:mean_square_displacement}
\end{figure}

\noindent Note that for large $J$ the contribution of higher modes decays only linearly instead of quadratically, demonstrating the relevance of the moment of inertia as an important control parameter.

Last, we address the mean-square displacement, which is most commonly investigated for passive and active Brownian motion. 
In Fig.~\ref{fig:mean_square_displacement}, we compare the experimentally determined mean-square displacement with the corresponding theoretical result. For short times, the particle is moving ballistically, as $\mean{\Delta \vec{r}^{2}(t)} = \mean{\dot{\vec{r}}^2(0)} \, t^2 + \mathcal{O}(t^3)$ (see Fig.~\ref{fig:mean_square_displacement}a). 
For larger times, the particle transitions towards a diffusive regime $\mean{\Delta \vec{r}^{2}(t)}\sim 4 D_\text{L} t$, which is characterized by the long-time diffusion coefficient 
\begin{equation}
	D_\text{L} = D_\text{t} + \sum_{k=1}^{\infty} \abs{\vec{c}_k}^{2} \Re \{  \mode_{k} \}. \label{eq:long_time_diffusion}
\end{equation}
Similar to the mean displacement, the mean-square displacement is affected by the confinement for long times which hinders the particle to reach a diffusive state.
In Fig.~\ref{fig:mean_square_displacement}b, we show the mean-square displacement parallel and perpendicular to the grooves comparing experiment, theory, and simulation. 
The mean-square displacement is non-monotonic in time due to the confinement.
At longer times, the particle needs to reorient before hitting the wall. The non-monotonic behavior results from the persistency of the particle and therefore is not observed for passive particles. 
The particle makes larger displacements along the grooves than perpendicular to them. 
In the absence of confinement, this anisotropy can persist even in the long-time limit characterized by the long-time diffusion matrix 
\begin{equation}%
	\big(\vec{D}_\text{L}\big)_{ij} = D_\text{t} \delta_{ij} +  \sum_{k=1}^{\infty}  \left( \text{c}_{k,i} \text{c}_{-k,j} + \text{c}_{-k,i} \text{c}_{k,j} \right) \Re\{ \mode_{k} \}, %
\end{equation}%
for $i,j \in \{x,y\}$. The eigenvalues of this matrix are given as $D_{\pm} = D_{ \text{L} } \pm \Del{D}_{ \text{L} }$, with the long-time anisotropy
\begin{align}%
	\Del{D}_\text{L} = \Big( \sum_{k,l=1}^{\infty} & \left( \abs{ \vec{c}_{k} \cdot \vec{c}_{l} }^{2} + \abs{ \vec{c}_{k} \cdot \vec{c}_{-l} }^{2} - \abs{ \vec{c}_{k} }^{2} \abs{ \vec{c}_{l} }^{2} \right) \nonumber  \\
	& \qquad \qquad \times \Re\{ \mode_k \} \Re\{ \mode_l \} \Big)^{1/2}, 
\end{align}%
which describes the long-time diffusion along the principal axes of maximal and minimal diffusion, respectively. 
The existence of a long-time anisotropy $\Del{D}_\text{L}  \neq 0$ will depend in general on the specific form of $\vec{v}(\phi)$.

\section{Discussion}

\begin{figure}
	\includegraphics[width=\columnwidth]{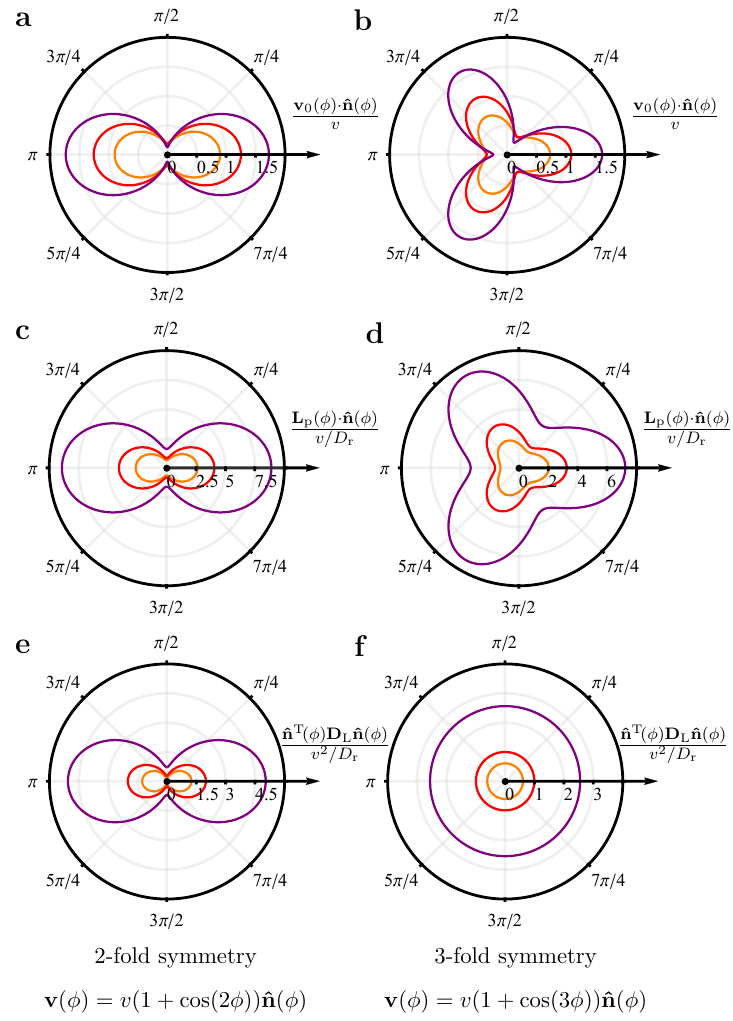}
	\caption{Anisotropy of the stationary mean velocity $\vec{v}_0$, persistence length $\vec{L}_\text{p}$, and long-time diffusion $\vec{D}_\text{L}$ for various values of the moment of inertia $J$ evaluated for a 2-fold symmetric motility (left column) and a 3-fold symmetric motility (right column). 
	\textbf{a}, \textbf{b} Stationary mean velocity as a function of the current orientation $\vec{v}_0(\phi) \cdot \uvec{n}(\phi)$.
	\textbf{c}, \textbf{d} Persistence length as a function of the initial orientation $\vec{L}_\text{p}(\phi) \cdot \uvec{n}(\phi)$.
	\textbf{e}, \textbf{f} Long-time diffusion projected along different directions $ \uvec{n}^\text{T}\!(\phi) \, \vec{D}_\text{L} \, \uvec{n}(\phi)$.
	The moment of inertia is set to $J = 0.1\,\gamma_\text{r}/D_\text{r}$ (orange), $J =  \gamma_\text{r}/D_\text{r}$ (red), and $J = 10\, \gamma_\text{r}/D_\text{r}$ (purple). The mass is fixed at $M= \gamma_\text{t}/ D_\text{r}$.}
	\label{fig:motility}
\end{figure}

Anisotropic motility has a strong impact on the motion of active particles both on short and long time scales. 
Our experiments demonstrate this explicitly for short and intermediate times and implicitly for long time-scales through simulations. Anisotropy persists for long times in the mean and mean-square displacement. 
We derived an analytical description that explains this behavior in terms of the Fourier series of the anisotropic driving term. 
The Fourier modes of the motility are linked to different time scales that add up and have an effect on the stationary mean velocity, persistence length and long-time diffusion.
Specifically, these quantities are mostly affected by the low-order Fourier coefficients. 

Our theoretical results predict that the degree of anisotropy is not only set by the orientation-dependent motility itself but depends non-trivially on all time scales $1/D_\text{r}$, $1/\abs{\omega}$, $\tau_\text{t}$, and $\tau_\text{r}$ of the model.
In Fig.~\ref{fig:motility}, we depict the anisotropy of the stationary mean velocity, persistence length, and long-time diffusion for different values of the moment of inertia $J$ and two exemplary orientation-dependent motilities $\vec{v}(\phi) = v (1 + \cos(n \phi)) \uvec{n}(\phi)$ with 2-fold symmetry ($n=2$) and 3-fold symmetry ($n=3$). 
In general, the mass and the moment of inertia have contrary effects on the anisotropy for short and intermediate times. 
For increasing mass, the dynamics of the particle involves stronger delay effects, smoothing the trajectories of the particle and effectively decreasing the anisotropy. 
On the other hand, increasing the moment of inertia leads to more resistance to reorientation and subsequently to higher persistence. 
The stationary parallel velocity in Fig.~\ref{fig:motility}a,b shows an increasing degree of anisotropy (being the ratio of outermost points to the innermost points on these curves) for increasing moment of inertia $J$. 
For the persistence length (see Fig.~\ref{fig:motility}c,d), the degree of anisotropy remains fairly invariant with increasing $J$ but overall we find a large persistence length (recalling Eq.~\eqref{eq:persistence times_asymptotic}).
Note that the mean displacement and thus the persistence length inherit the symmetry of the driving velocity $\vec{v}(\phi)$. 
This symmetry is in general lost for long times, since the long-time diffusion can either follow a 2-fold symmetric modulation or behaves fully isotropic in every direction (see Fig.~\ref{fig:motility}e,f). 
In fact, for motilities with higher rotational symmetry than two-fold, the long-time diffusion is always isotropic. 
Thus, we like to stress that even a system showing isotropic diffusion can hide anisotropic dynamics on shorter time scales.  

\begin{figure}
	\includegraphics[width=\columnwidth]{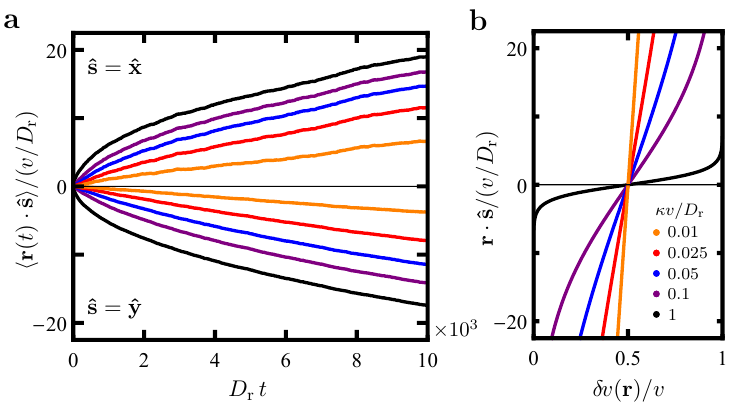}
	\caption{Mean displacement of a particle in an anisotropy gradient. \textbf{a} Mean position in the gradient direction, $\mean{\vec{r}(t) \cdot \uvec{s}}$ for different reduced growth rates $\kappa v/ D_\text{r}$. The upper plane displays simulated data for the gradient direction $\uvec{s} = \uvec{x}$, while the lower plane shows the data for $\uvec{s} = \uvec{y}$.
    \textbf{b} The anisotropy $\del{v}(\vec{r})$ of the orientation-dependent motility is plotted as a function of position $\vec{r}\cdot \uvec{s}$, using the same reduced growth rates as in \textbf{a}. The plot is sideways to align with \textbf{a}.}
	\label{fig:anisotropic_gradients}
\end{figure}

As an outlook, we want to highlight the intriguing possibilities that arise from combining position- and orientation-dependent motility. 
This opens up avenues to explore migration in gradients of anisotropy. Our experimental system relies on pre-molded lenticular sheets to create the anisotropic substrate. However, by employing a larger 3D printer or an engraving tool, more complex substrates could be generated, for instance, to introduce gradients in anisotropy.
For proof of principle, we assume that the particle exhibits 2-fold symmetric motility $\vec{v}(\vec{r},\phi) = \left( v+ \del{v}(\vec{r}) \cos(2 \phi) \right) \uvec{n}(\phi)$, where the anisotropy $\del{v}(\vec{r})$ increases along the direction of $\uvec{s}$. 
Specifically, we employ a logistic function to describe the spatial dependency, $\delta v(\vec{r}) = v/(1 + e^{-\kappa \vec{r} \cdot \uvec{s}})$, with $\kappa$ representing the growth rate. This function yields maximum anisotropy ($\del{v}(\vec{r}) = v$) for $\vec{r} \cdot \uvec{s} \to \infty$, and isotropic motility ($\del{v}(\vec{r}) = 0$) for $\vec{r} \cdot \uvec{s} \to - \infty$, with a symmetric slope around the origin (see Fig.~\ref{fig:anisotropic_gradients}b).
For simplicity, we consider only the overdamped case ($m=J=0$) and examine the mean position along the gradient $\mean{ \vec{r} (t) \cdot \uvec{s}}$  for particles initially starting in the origin (which corresponds to the inflection point of $\del{v}(\vec{r})$).
In Fig.~\ref{fig:anisotropic_gradients}a, we present the mean position for different growth rates $\kappa$ and gradient directions  $\uvec{s}$. 
We observe opposing behavior depending on whether the gradient is aligned with $\uvec{x}$ or $\uvec{y}$, \ie, positive displacement when $\uvec{s}=\uvec{x}$ and negative displacement for $\uvec{s}=\uvec{y}$.
Thus, the particle exhibits motion parallel or antiparallel to the gradient towards regions where the motility increases. Specifically, for $\uvec{s}=\uvec{x}$ the particle moves towards more anisotropic regions, whereas for $\uvec{s}=\uvec{y}$ it moves towards more isotropic regions.
Since the rotational dynamics is independent of the particle position, there is always an equal probability of moving upward and downward the gradient. However, when particles move towards regions of higher motility, they experience simultaneous acceleration within the persistence time. Consequently, the persistence length is always greater in the direction of increasing motility compared to the opposite direction, indicating a displacement for long times towards regions of higher motility.
We like to stress that this result is not contradictory to previous studies in spatial motility fields, where the stationary positional probability shows accumulation in regions of low motility \cite{Schnitzer1993}.  
Here, we are considering a gradient in an unbounded space. Thus, we are not reaching stationarity within finite simulation time.

Our model could be useful to predictively optimize driving parameters for the navigation of active matter in anisotropic environments \cite{SelmkeKBCY2018a,BreoniSL2020,RodriguezGARBVI2020,DaddiMoussaIderLL2021}, for instance robotic systems. 
In particular, the persistence length is an important control parameter that strongly impacts collective phenomena, like motility-induced phase separation \cite{ButtinoniBKLBS2013,CatesT2015,CapriniGL2022}. Swarms of self-propelled particles moving with an orientation-dependent motility would be an interesting topic for future research, for which our model provides a baseline \cite{CohenG2014,LavergneWBB2019,CasiulisTCD2020,SolonCTT2022}.

\section{Methods}

\subsection{Particle fabrication} 

The particle used in this work has been manufactured by 3D-printing using a stereolithographic acrylic based photopolymer 3D printer (Formlabs Form 2, using Grey V3 material, identical to Ref.~\cite{ScholzJLL2018}).
Figure~\ref{fig:experiment}a shows an image of the particle. It consists of a cylindrical core (diameter 9~mm, height 4~mm) and a cap (diameter 15~mm, height 2~mm).
Seven tilted cylindrical legs (diameter 0.8~mm, inclination angle 4 degrees) are attached to the cap in a regular heptagon around the bottom cylinder. 
The legs are tilted parallel to each other defining the orientation of the particle. 
The length of the legs is chosen such that the bottom of the particle is lifted by 1 mm above the surface. 
The particle is marked with a sticker from which the orientation can be determined using computational image processing. The particle's mass is about $m=0.76\,\text{g}$. From the particle's mass and shape, its moment of inertia is computed to be $J=1.64\times10^{-8}\,\text{kg}~\text{m}^2$, assuming homogeneous density.\\

\subsection{Experimental setup and analysis} 

Particle motion is excited by vertical vibrations of a rectangular acrylic baseplate (side length 300~mm, thickness 15~mm) with a lenticular plastic sheet on top, attached to an electromagnetic shaker (Tira TV 51140). The sheet's surface consists of equally spaced elliptical half-cylinders with a density of $0.787~\text{mm}^{-1}$ (20 lines per inch) and a groove depth of 0.315 mm. An illustration and a cross-section of the particle resting on such a grooved surface are shown in Fig.~\ref{fig:experiment}c, respectively. Lenticular sheets of this kind are typically used in digital printing or displays to create images with the illusion of depth. Here, we use it to induce an anisotropic driving of the particle parallel and perpendicular to the lines, since the speed of the particle is very sensitive to the contact angle of the legs to the surface. Note that the width and height of the grooves are chosen such that the particle legs cannot be significantly trapped (see Fig.~\ref{fig:experiment}c), in order to prevent the particle simply from sliding along grooves.

The tilt of the plate is adjusted with an accuracy of $0.01^{\circ}$ to minimize gravitational drift. The vibration frequency is set to $f=80\,\mathrm{Hz}$ and three different peak acceleration amplitudes $A = 1.28\,g$, $1.44\,g$ and $1.60\,g$ are studied.

A mid-to-high-speed camera system (Allied Vision Mako-U130B) operating at 150 frames per second is used to record the experiment with a spatial resolution of $1024 \times 1024$ pixels. The particle location and orientation are determined and tracked using standard image recognition methods (Hough transform and morphological image region analysis) to a spatial accuracy of about $\pm 3\times10^{-5}\,\text{m}$ and a orientational accuracy of $\pm0.74^{\circ}$ \cite{ScholzJLL2018}. Multiple single trajectories are recorded for each amplitude, until 20 min of data are acquired per recording. Half of the recorded time the particle starts parallel and the other half of the time it starts perpendicular to the grooves. Events involving particle-border collisions mark a trajectory's termination and are subsequently discarded, resulting in trajectories of various lengths.

The velocity was calculated from the displacement of successive positions of the particle as $\vec{v}(t) = \left( \vec{r}(t +\Del{t}) - \vec{r}(t) \right)/\Del{t}$, where $\Del{t} = 1/150\,\text{s}$ is the time between two frames. The time steps are not fully equidistant between recorded frames, therefore the experimental data were linearly interpolated to obtain equidistant points. 
Experimental means with respect to a specific initial orientation $\phi_{0}$ were calculated by averaging in the interval $[\phi_{0} - \del{\phi}, \phi_{0} + \del{\phi}]$. We chose $\del{\phi} = 10^{\circ}$ and modified the theoretical results accordingly by $\exp(\ii k \phi ) \to \exp(\ii k \phi ) \sin(k \del{\phi}) /(k \del{\phi})  $. We took advantage of the rotational and inflection symmetries of the experiment (by rotating some trajectories by 180 degrees) to increase the angular statistics for the mean displacement. \\

\begin{figure*}
	\includegraphics[width=17cm]{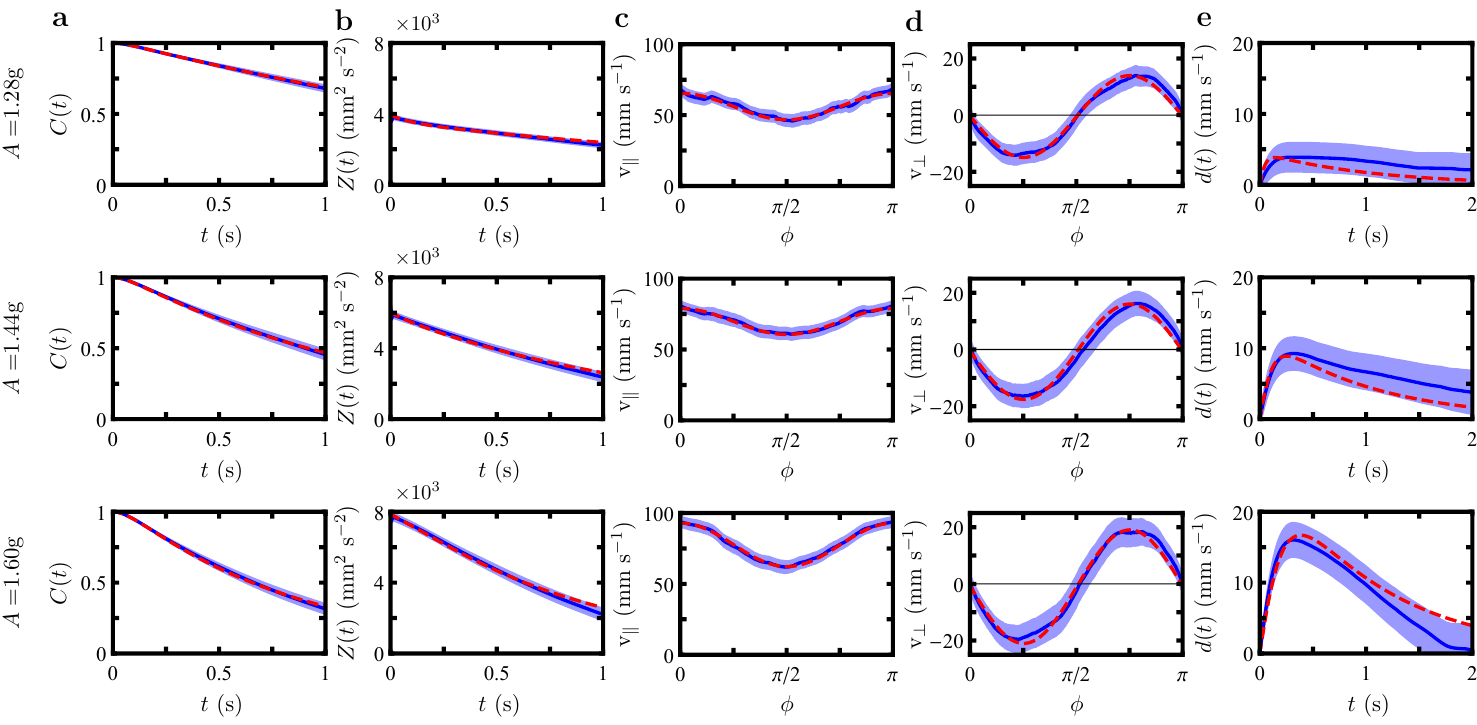}
	\caption{Determination of model parameters. \textbf{a} Orientational correlation function $\fct{C}(t)$, \textbf{b} velocity correlation function $\fct{Z}(t)$, \textbf{c} stationary parallel velocity $\text{v}_\parallel$ \textbf{d} stationary perpendicular velocity $\text{v}_\perp$. Solid dark blue and dashed red curves show the experimental data and analytical results, respectively. Experimental error intervals represent the standard error of the mean. The parameter values are listed in Tab.~\ref{table:parameter}. \textbf{e} Time-dependence of the delay function $\fct{d}(t)$ validating the parameters on an independent quantity. The different vibration amplitudes are A = 1.28 g (upper row), A = 1.44 g (middle row), and A = 1.60 g (lower row).}
	\label{fig:parameter}
\end{figure*}

\subsection{Analytic results}

Both the translational velocity $\dot{\vec{r}}(t)$ and the angular velocity $\dot{\phi}(t)$ undergo a simple stochastic process for which a general solution is easily obtained (see Eqs.~\eqref{eq:langevin_r} and \eqref{eq:langevin_phi}). 
Several dynamical correlation function as well as low-order moments can be consequently calculated using standard methods of stochastic calculus \cite{Risken1996}. 
The orientational correlation function $\fct{C}(t) = \mean{\uvec{n}(t) \cdot \uvec{n}(0)}$ displays a double exponential decay 
\begin{equation}
\fct{C}(t)  = \cos(\omega t) e^{- D_\text{r} \left( t - \tau_\text{r} \left( 1 - e^{-t/\tau_\text{r}} \right) \right)},
\label{eq:orientational_correlation}%
\end{equation}
(as previously discussed in Ref.~\cite{GhoshLMM2015,WalshWSOBM2017,ScholzJLL2018}). 
The velocity correlation function $\fct{Z}(t) = \mean{\dot{\vec{r}}(t) \cdot \dot{\vec{r}}(0)}$ is given as
\begin{equation}
\fct{Z}(t)  =2 \frac{D_\text{t}}{\tau_\text{t}} e^{-t/\tau_\text{t}} + 2 \sum_{k=1}^{\infty} \abs{\vec{c}_{k}}^2 \Re \{ \fct{V}_{k}^{+}(t) \},
\label{eq:velocity_correlation}%
\end{equation}
where the Fourier-coefficient vectors are determined by the orientation-dependent motility, as $\vec{c}_k = \integral{-\pi}{\pi}{\vec{v}(\phi) \exp(- \ii k \phi) /(2 \pi) }{\phi}$  (see Eq.~\eqref{eq:velocity}), and
{\allowdisplaybreaks
	\begin{align}
		& \fct{V}^{\pm}_{k}(t)  =  \frac{\tau_\text{r}}{\tau_\text{t}} \frac{e^{ \text{S}_{k} }}{2} \bigg( \pm \text{S}_{k}^{-\Omega_{k}^{+} } \Gamma\left(\Omega_{k}^{+},0, \text{S}_{k}e^{-t/\tau_\text{r}}\right) e^{t/\tau_\text{t}} \nonumber \\
		& \hspace{2.25cm} - \text{S}_{k}^{-\Omega_{k}^{-} } \Gamma\left(\Omega_{k}^{-},0, \text{S}_{k}e^{-t/\tau_\text{r}}\right) e^{-t/\tau_\text{t}} \\
		& + \left( \text{S}_{k}^{-\Omega_{k}^{+} } \Gamma\left(\Omega_{k}^{+},0, \text{S}_{k}\right) + \text{S}_{k}^{-\Omega_{k}^{-} } \Gamma\left(\Omega_{k}^{-},0, \text{S}_{k}\right) \right) e^{-t/\tau_\text{t}}  \bigg), \nonumber
\end{align}}%
with  $\Omega_{k}^{\pm} = D_\text{r} \tau_\text{r} k^2 \pm( \ii \omega \tau_\text{r} k + \tau_\text{r} / \tau_\text{t}) $ and $\text{S}_k =  D_\text{r} \tau_\text{r} k^2$. The real part is denoted by $\Re \{ \dots \}$ and the generalized incomplete gamma function is  $\Gamma(s,x_{1},x_{2}) = \integral{x_{1}}{x_{2}}{ t^{s-1} e^{-t}}{t}$. 
The delay function measuring the difference between the direction of the velocity and the current orientation, $\fct{d}(t) = \mean{\dot{\vec{r}}(t)\cdot\uvec{n}(0)} - \mean{\dot{\vec{r}}(0)\cdot\uvec{n}(t)}$, is given by 
{\allowdisplaybreaks
	\begin{align}\label{eq:delay_function}
		& \fct{d}(t)  = \Re \{  \big( \text{c}_{1,x} + \text{c}^{*}_{1,x} + \ii ( \text{c}_{1,y} - \text{c}^{*}_{1,y})  \big) \fct{V}^{-}_{1}(t) \}, 
\end{align}}%
which coincides with the result for isotropic self-propulsion \cite{ScholzJLL2018} (due to the projection onto the orientation).
Next, we give the mean displacement $\mean{\Delta \vec{r}(t)} = \mean{\vec{r}(t) - \vec{r}_0}$ under the condition that initially the position $\vec{r}_0 $ and the orientation $\phi_{0}$ are prescribed,
\begin{equation}
\mean{\Delta \vec{r}(t)} = \vec{v}_{0} \tau_\text{t} (1-e^{-t/\tau_\text{t}}) +  \sum_{\substack{k=-\infty \\ k\neq 0}}^{\infty} \! \vec{c}_{k} \fct{R}_{k}(t) e^{\ii k \phi_{0}}, 
\label{eq:mean_displacement}%
\end{equation}
with the stationary velocity $\vec{v}_{0}$ (see Eq.~\eqref{eq:stationary_velocity}),
{\allowdisplaybreaks
	\begin{align}
	\fct{R}_{k}(t)  = & \tau_\text{r} e^{ \text{S}_{k} } \bigg( \text{S}_{k}^{-\Omega_{k}} \Gamma\left(\Omega_{k}, \text{S}_{k}e^{-t/\tau_\text{r}},\text{S}_{k}\right)  \\
	& - \text{S}_{k}^{-\Omega_{k}^{-} } \Gamma\left(\Omega_{k}^{-}, \text{S}_{k}e^{-t/\tau_\text{r}},\text{S}_k\right) e^{-t/\tau_\text{t}} \bigg), \nonumber
	\label{}%
	\end{align}}
\noindent and $\Omega_{k} = D_\text{r} \tau_\text{r} k^2 + \ii \omega \tau_\text{r} k$ .
Lastly, we provide the result for the mean-square displacement $\mean{\Delta \vec{r}^{2}(t)} = 	\mean{(\vec{r}(t) -\vec{r}_0)^{2} }$ which can be expressed as 
\begin{equation}
	\mean{\Delta \vec{r}^{2}(t)} = 4 D_\text{L} t + 2 \big( \fct{Z}(t) - \fct{Z}(0) \big) \tau_\text{t}^{2} - 4 \fct{F}(t) \tau_\text{r}^{2}
	\label{eq:mean_square_displacement}%
\end{equation}
with the long-time diffusion coefficient $D_\text{L}$ (see Eq.~\eqref{eq:long_time_diffusion}), the velocity correlation function $\fct{Z}(t)$ (see Eq.~\eqref{eq:velocity_correlation}) and
{\allowdisplaybreaks
	\begin{align}
	\fct{F}(t)  = &  \sum_{k=1}^{\infty} \abs{\vec{c}_k}^{2} \Re \Bigg\{ \frac{e^{ \text{S}_{k} }}{\Omega_{k}^2} \Bigg(  \pFq{2}{2}{\Omega_{k},\Omega_{k}}{\Omega_{k}+1,\Omega_{k}+1}{-\text{S}_k}  \\
	&-  \pFq{2}{2}{\Omega_{k},\Omega_{k}}{\Omega_{k}+1,\Omega_{k}+1}{-\text{S}_k e^{-t /\tau_\text{r}}} e^{-\Omega_{k} t /\tau_\text{r}}  \Bigg) \Bigg\}, \nonumber
	\end{align}}
\noindent where ${}_{2}F_{2}$ denotes the generalized hypergeometric function.
Last we remark that in the overdamped limit, \ie~$m\to0$ and $J\to0$, we recover the results of orientation-dependent motility in underdamped systems \cite{SprengerFRAIWL2020} and similarly for an isotropic self-propulsion $\vec{v}(\phi)= v_0 \uvec{n}(\phi)$, we obtain the expressions of Ref.~\cite{SprengerJIL2021}.

\subsection{Parameter estimation}

The underdamped active Brownian motion model depends on eight independent parameters. All parameters were obtained using the MATLAB standard optimizer $\texttt{fminsearch}$ (Nelder-Mead optimization of a function of several variables on an unbounded domain). Our cost function consists of five terms covering different parameters. Each term is constructed as follows: The absolute deviation between the experimental mean and the analytical expectation is weighted with the standard error of the mean and then averaged over time or orientation. This procedure takes into account the experimental uncertainty \cite{BaileySGLI2022}. At the same time, the value of our cost function quantifies the fit itself. We call a fit sufficiently representative of the experimental mean if the mean deviation between experimental mean and analytical expectation is no greater than one standard error. We use this definition to determine an error interval for our optimal parameters. 
The orientational correlation function $\fct{C}(t)$ (see Eq.~\eqref{eq:orientational_correlation}) is used to determine the rotational diffusion constant $D_\text{r}$ and the rotational friction time $\tau_\text{r}$.
In addition, we use the mean stationary angular velocity $\mean{\dot{\phi}(0)} = \omega$ to determine the angular speed $\omega$. Further, we use the velocity correlation function $\fct{Z}(t)$ (see Eq.~\eqref{eq:velocity_correlation}) to extract values for the translational friction time $\tau_\text{t}$ and the translational short-time diffusion coefficient $D_\text{t}$.
Lastly, we use the mean stationary velocity $\vec{v}_{0}$ (see Eq.~\eqref{eq:stationary_velocity}), which is projected parallel ($\text{v}_\parallel = \vec{v}_{0} \cdot \uvec{n}$) and perpendicular ($\text{v}_\perp = \vec{v}_{0} \cdot \uvec{n}_\perp$) to the body axis, to determine all the motility parameters $\text{v}_{\perp}$, $\del{\text{v}}_{\parallel}$, and $\del{\text{v}}_{\perp}$.
In Fig.~\ref{fig:parameter}a-d, the analytic fitting curves to the experimental data are shown and the resulting set of parameter is listed in Tab.~\ref{table:parameter}. For vibrobots, the delay function $\fct{d}(t)$ (see Eq.~\eqref{eq:delay_function}) proved to be a sensitive measure for the quality of the determined parameter-set \cite{ScholzJLL2018}. Figure~\ref{fig:parameter}e shows good agreement between theory and experiment for all three measurements. Note that $\omega$ and $D_\text{t}$ are not significantly different from zero for our particles. However, they are included in the model, since they can be relevant for different experimental realizations in the literature \cite{ScholzJLL2018,DauchotD2019, Siebers2023}. For the inertial time scales $\tau_\text{r}$ and $\tau_\text{t}$ we see an increase for increasing $A$, and $\tau_\text{t}$ only becomes significant for $A=1.44\,g$ and $A=1.60\,g$. This is likely caused by the reduction of friction at larger $A$.

\begin{table} [h]
	\caption{\label{table:parameter} Model parameters obtained from analytical fits to measurements in Fig.~\ref{fig:parameter}. Lower and upper 95\% confidence bounds are displayed behind each value. Parameters marked by $^*$ are not significantly different from zero.}
	\begin{ruledtabular}
		\begin{tabular}{l@{\hspace{-0cm}}l@{\hspace{0.7cm}}ccc}
			A & (g) & 1.28 & 1.44 & 1.60 \\
			\hline
			$\omega^*$ &  ($\text{s}^{-1}$) & 0.09 $\substack{+0.76 \\ -0.76}$ & 0.12 $\substack{+0.88 \\ -0.99}$ & 0.11 $\substack{+0.89 \\ -1.11}$ \\
			$D_\text{r}$ & ($\text{s}^{-1}$) & 0.39 $\substack{+0.05 \\ -0.04}$ & 0.80 $\substack{+0.07 \\ -0.10}$ & 1.18 $\substack{+0.11 \\ -0.13}$ \\
			$\tau_\text{r}$ & (s) & 0.05 $\substack{+0.02 \\ -0.02}$ & 0.06 $\substack{+0.03 \\ -0.01}$ & 0.07  $\substack{+0.03 \\ -0.02}$ \\
			$\text{v}_{\parallel}$ & (mm $\text{s}^{-1}$) & 57.5 $\substack{+4.8 \\ -4.1}$ & 73.2 $\substack{+4.9 \\ -4.4}$ & 85.0  $\substack{+5.3 \\ -4.8}$ \\
			$\del{\text{v}}_{\parallel}$ & (mm $\text{s}^{-1}$) & 9.2 $\substack{+7.0 \\ -6.8}$ & 8.5 $\substack{+7.5 \\ -7.3}$ & 15.7 $\substack{+8.6 \\ -8.5}$ \\
			$\del{\text{v}}_{\perp}$ & (mm $\text{s}^{-1}$) & 15.6 $\substack{+5.7 \\ -5.5}$ & 19.3 $\substack{+7.1 \\ -7.0}$ & 23.8  $\substack{+9.4 \\ -9.1}$ \\ 
			${D_\text{t}}^*$ & (mm$^2$ $\text{s}^{-1}$) & 27.89 $\substack{+31.85 \\ -27.89}$ & 36.23 $\substack{+44.38 \\ -36.23}$ & 59.41  $\substack{+40.59 \\ -59.41}$ \\
			$\tau_\text{t}$ & (s) & 0.07 $\substack{+0.10 \\ -0.07}$ & 0.10 $\substack{+0.09 \\ -0.06}$ & 0.13  $\substack{+0.06 \\ -0.06}$ \\
		\end{tabular}
	\end{ruledtabular}
\end{table}

\subsection{Simulation}

Numerical data for a self-propelled particle with orientation-dependent motility enclosed by absorbing boundaries are included in Figs.~\ref{fig:anisotropic_delay}, \ref{fig:mean_displacement}b, and \ref{fig:mean_square_displacement}b. Equations~\eqref{eq:langevin_r} and \eqref{eq:langevin_phi} were discretized to perform Langevin dynamics simulations using first-order finite difference discretization. For these simulations, we chose the time step size $\Delta t=10^{-2}$~s and we performed $10^{5}$ realizations in Fig.~\ref{fig:mean_displacement}b, and \ref{fig:mean_square_displacement}b and $2000$ realizations in Fig.~\ref{fig:anisotropic_delay} to calculate the respective ensemble averages. 
Half of the trajectories started at $x_0 = 0 \,\text{mm}$, $y_0 = -100 \,\text{mm}$, and $\phi_0 = \pi/2$ and the other half at $x_0 = 100 \,\text{mm}$, $y_0 = 0 \,\text{mm}$ and $\phi_0 = \pi$ (modelling the initialisation in the experiment).
The rectangular absorbing boundary was set at $\{ (x,y) | (x = \pm 130 \, \text{mm},\,y\in [-130 \, \text{mm}, \, 130 \, \text{mm}])   \lor (x\in [-130 \, \text{mm}, \, 130 \, \text{mm}],\, y = \pm 130 \,\text{mm}) \}$. 

Figure~\ref{fig:anisotropic_gradients} presents simulation data for a self-propelled particle with additional positional dependency in its motility $\vec{v}(\vec{r},\phi)$. In this case, we used a time-step size of $\Delta t = 10^{-1}$~s and performed $10^{5}$ realizations to compute the ensemble averages. Each trajectory started from the origin $(x_0, y_0) = (0, 0)$ with a random orientation.

\subsection{Orientation-dependent friction and torque}

\begin{figure}
	\includegraphics[width=\columnwidth]{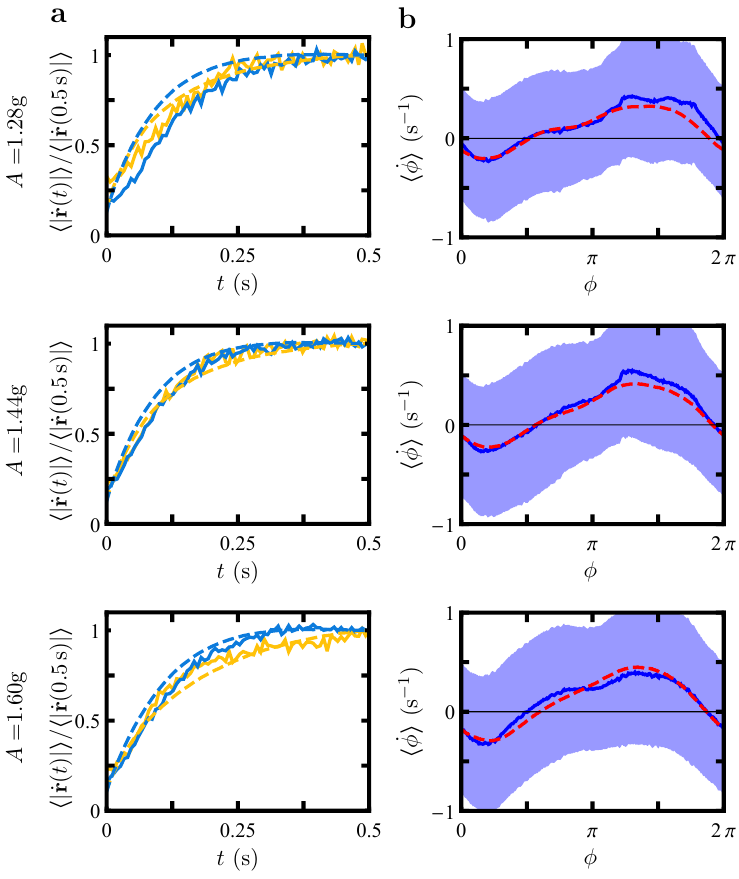} 
	\caption{Initial increase of the velocity and orientation-dependence of the angular velocity. 
    \textbf{a} Normalized initial velocity $\bm{\langle} \abs{\dot{\vec{r}}(t)}\bm{\rangle} / \bm{\langle} \abs{\dot{\vec{r}}(0.5\,\text{s})}\bm{\rangle}$ of a particle starting from rest (solid curves) for initial orientations $\phi_0 = 0$ (cyan) and $\phi_0 = \pi/2$ (yellow). The difference between horizontal and vertical starting orientation is small and within the variation observed in simulations (dashed lines). This suggests that friction in the parallel and perpendicular direction is not significantly different. 
    \textbf{b} Mean angular velocity $\mean{\dot{\phi}}$ as a function of particle orientation $\phi$ (solid curves). The slight periodicity is an artifact of the initial conditions and can be reproduced by simulating with equal conditions (dashed curves).
	Both are shown for excitation amplitudes $A$ = 1.28\,g (upper row), $A$ = 1.44\,g (middle row), and $A$ = 1.60\,g (lower row).}
	\label{fig:torque_friction_anisotropy}
\end{figure}

In our underdamped active Brownian particle model, the influence of an anisotropic environment is effectively described by an orientation-dependent motility. However, in more general cases, self-propelled particles may undergo more complicated dynamics in anisotropic environments, resulting in orientational dependencies in various model parameters beyond motility. Here, we provide supplementary information on why we chose not to model anisotropic motion using an anisotropic friction matrix or orientation-dependent torque.

While using an anisotropic friction matrix may seem like an intuitive approach to describe anisotropic motion, it is not suitable for our experimental particles. Most of the time, our particles do not have direct contact with the substrate, and energy dissipation only occurs during collisions. Only the effective angle between legs and plate is different in perpendicular and parallel directions causing anisotropic self-propulsion. Experimental evidence supporting this conjecture is presented in Fig.~\ref{fig:torque_friction_anisotropy}a, showing the initial increase of the velocity, which reaches an intermediate state after approximately $0.5$~s regardless of the initial orientation. From this we conclude an isotropic translational damping time and thus isotropic friction.

Furthermore, particles may experience anisotropic torques. Surprisingly, measuring the angular velocity for given orientations suggests no such torques in our experiment (see Fig.~\ref{fig:torque_friction_anisotropy}b). The slight modulation observed in the angular velocity, which does not align with the overall two-fold symmetry of the environment, can be attributed to initialization bias. Simulation results assuming isotropic torque, indicated by the black curves in Figure 8b, are consistent with the experiment.

\bigskip

\section{Data availability}
Illustrative videos of the experiments are available as Supplementary Movies 1-–6.
Raw data supporting the results of this work are available at https://doi.org/10.5281/zenodo.7220326.

\section{Code availability}
All custom simulation and analysis code used to derive the results presented herein is available at https://doi.org/10.5281/zenodo.7220326.

\section{Acknowledgments}  
C.S., R.W., and H.L.\ are funded by the Deutsche Forschungsgemeinschaft (DFG, German Research Foundation) -- SCHO 1700/1-1; 283183152 (WI 4170/3-2); LO 418/23-1.

\section{Author contributions}  
C.S. and H.L. designed the research. C.S. designed the experimental setup. C.S. and A.L. carried out the experiments. A.S. and C.S. analyzed the measurements. A.S. developed the theoretical and numerical results. All authors discussed the results and wrote the manuscript.

\section{Competing interests}  
The authors declare no competing interests.

\section{Additional information}
Correspondence and requests for materials should be addressed to A.S. (email:~alexander.sprenger@hhu.de)  or to C.S. (email:~christian.scholz@hhu.de).

\nocite{apsrev41Control}
\bibliographystyle{apsrev4-1}
\bibliography{control,refs}

\end{document}